\documentclass[12pt]{article}
\usepackage[utf8]{inputenc}
\usepackage[T1]{fontenc}
\usepackage{authblk}
\usepackage{amsmath,amssymb}
\usepackage{graphicx}
\usepackage{hyperref}
\usepackage{times}
\usepackage{cite}
\usepackage{tabularx}
\usepackage{booktabs}
\usepackage[utf8]{inputenc}
\title{Zeno's paradox and black hole information loss problem}
\author[1]{Xian-Hui Ge}
\affil[1]{Department of Physics, College of Siences, Shanghai University, Shanghai 200444, China}
\date{}

\begin{document}
\maketitle

\begin{abstract}
We develop a conceptual parallel between the black hole information problem and Zeno's paradox, highlighting the role of limiting procedures that turn formally infinite constructions into finite physical observables. Building on the replica--wormhole paradigm, we move beyond unitarity restoration to formulate a quantitative notion of irreversibility in Hawking radiation. Our main result is a modular thermodynamic framework for black-hole evaporation, in which modular entropy, entanglement capacity, and relative entropy assume thermodynamic roles. The monotonicity of relative entropy furnishes a generalized second law that determines the arrow of evolution in replica space. We further resolve the apparent tension between the replica method and the quantum no-cloning theorem by interpreting replicas as ensemble representations rather than physical copies of an unknown state, thereby clarifying the operational meaning of gravitational path integrals. A key message of this work is that non-additivity in Tsallis statistics provides an information-theoretic analogue of the correlations induced by replica wormholes.
\end{abstract}


\section{Introduction}

\quad The black hole information loss problem, known as the black hole information paradox, refers to the fact that semiclassical calculations based on quantum fields in curved spacetime show that the black hole evaporation process is not unitary, but the holographic principle and string theory imply that this process should be unitary~\cite{1Hawking1975,2Hawking1976}. Specifically, the Hawking radiation spectrum shows that the black hole evaporation process follows an equilibrium blackbody spectrum, which means that information seems to be permanently lost in the black hole evaporation process, and the information of the initial state cannot be recovered by observing the radiation. This non-unitarity conflicts with the basic principle of quantum mechanics, which requires that physical processes must be unitary, that is, information must be conserved. However, advances in string theory, especially in gauge/gravity duals such as AdS/CFT duality, provide an alternative perspective~\cite{3Susskind1993,4Susskind1994,5Maldacena1999}. These theories show that the formation and evolution of black holes are actually unitary. This is because physical processes in gauge field theory always satisfy unitarity, that is, information is not lost during evolution. Through gauge/gravity duality, the black hole problem in gravity theory can be mapped to a gauge field theory on the boundary, and in gauge field theory, unitary evolution is natural. Therefore, the core of the black hole information paradox lies in the contradiction between semiclassical computation and quantum gravity theory.  However, astrophysical black holes (such as Kerr black holes) exist in asymptotically flat spacetime, and their asymptotic structure is essentially different from that of AdS spacetime, which makes AdS/CFT impossible to apply directly. To deal with this problem in flat spacetime, we need to consider several aspects: the island mechanism ensures the unitary evolution of entropy by associating the internal information of the black hole with the external radiation through the emergence of quantum extremal surfaces. Recent studies have shown that in asymptotically flat two-dimensional gravity models (such as the JT gravity model), the island region dominates the calculation of entropy after Page time, and its behavior is similar to that of AdS black hole ~\cite{35Penington2022,36Almheiri2020}. The key elements of this mechanism may be independent of the AdS boundary condition~\cite{6Hartman2020,7Hashimoto2020,8Wang2021,9Kim2021,10Ling2021,11Yu2023,12Yu2022}. The author's early work on the information loss problem of black holes can be found in~\cite{13GeDissertation,14Ge2005a,15Ge2005b,16Ge2004a,17Ge2004b}.

In recent years, remarkable progress has been made in solving the problem of black hole information loss~\cite{18Page1993a,19Page2013,20Page1993b,21Lewkowycz2013,22Engelhardt2015,23Almheiri2019,24Almheiri2019b,25Penington2020,26Almheiri2021,27Almheiri2019c,28Hollowood2020,29Goto2021,30Miao2024,31Chu2021,32Chen2020,33Du2023,34Ling2022}. By developing the path integral method based on two-dimensional gravity and using the replica trick to calculate the R\'enyi entropy replica wormholes saddle, it is found that there is another saddle point solution based on the original Hawking's calculation, that is, the replica wormholes saddle. Then, by taking the limit back to the von Neumann entropy of Hawking radiation, one can give a Page curve of information conservation, i.e., the entropy of Hawking radiation for both the initial and final states is zero. This result shows that the process of black hole evaporation meets the requirement of unitarity of quantum mechanics, so there is no problem of information loss in the process of black hole evaporation. The related work is considered to be a key step towards solving the problem of black hole information loss, especially the development of two-dimensional gravitational path integral and the discovery of new saddle point solutions, which enrich the study of quantum gravity and become a hot topic in theoretical physics in recent years~\cite{35Penington2022,36Almheiri2020}.

This paper first reviews another ancient paradox, Zeno's paradox, and provides new insights into the problem of black hole information loss by analyzing the similarities between Zeno's paradox and the problem of black hole information loss. If the von Neumann entropy of Hawking radiation is regarded as a conserved quantity in the process of black hole evaporation, how to define a physical quantity to describe the ``change'' of Hawking radiation entropy?

Zeno, an ancient Greek philosopher and a staunch defender of the Parmenides school, is famous for such paradoxes as ``the flying arrow does not move'', ``Achilles and the tortoise'', and ``the dichotomy''. In the following, we will mainly focus on the proposition of ``the flying arrow does not move'', which is expressed as follows: Suppose a flying arrow is at a certain point in space at a certain time. At this particular moment, the arrow occupies a definite position. Since any object can only occupy one position at any instant, the arrow is stationary at that instant. It is further inferred that if the arrow is stationary at every instant, then the arrow is stationary throughout the flight. Therefore, it is logical to assume that the arrow does not actually move, but in fact the observer can see that the arrow does move, so it is called the ``flying arrow does not move'' paradox.

The solution of the paradox requires the introduction of the concepts of ``limit'' and ``instantaneous velocity''. As a mathematical concept, limit can be obtained from continuous motion by solving the limit, while the concept of instantaneous velocity is defined in the ``velocity space'' and does not exist in the physical space where the flying arrow is located. It exists in a completely different, abstract space: the tangent space of the flight vector trajectory. The concept of limit and its associated differential is the content of Newtonian particle mechanics. In addition to the concept of instantaneous velocity, Newton also introduced a more abstract concept: ``acceleration'', and considered that ``acceleration'' and velocity were two completely different concepts. It is on the basis of these concepts that the whole framework of Newtonian mechanics is established, which shows how deeply Zeno thought about the relationship between time, space and motion.

To motivate the role of limiting procedures in the black hole information
problem, it is useful to recall the conceptual structure behind Zeno's paradox.
Zeno's argument highlights the difficulty of describing continuous motion when
one attempts to reconstruct it solely from discrete snapshots. In modern physics
motion is taken as a basic physical fact, while differential calculus provides a
consistent framework to quantify it by a continuum limit,
\begin{equation}
v = \lim_{\Delta t \to 0} \frac{\Delta x}{\Delta t},
\end{equation}
which defines a tangent vector along an already existing trajectory. The paradox
is resolved not only by altering the underlying physics but also by adopting an appropriate
mathematical language. A formally similar structure appears in quantum information. The von Neumann
entropy arises as the continuum limit of the R\'{e}nyi family,
\begin{equation}
S_{\mathrm{vN}}(\rho)
= \lim_{n\to1} S_{n}(\rho),
\end{equation}
where the replica index $n$ is analytically continued from integer values to the
real line. This limiting procedure defines a consistent information measure, and
its utility becomes evident in gravitational systems, where replica geometries
and quantum extremal surfaces yield the Page curve. The analogy between Zeno limits and replica limits is therefore methodological
rather than mechanistic: in both cases a continuum limit organizes the correct
description, but only in the black hole setting do new gravitational saddles
(replica wormholes) enter and change the physical content of the theory.

Generalized entropies provide an additional effective language for
non-additive correlations. For example, the quantum Tsallis entropy \cite{tsallis,tsallis1},
\begin{equation}
S_{q}(\rho)
= \frac{1 - \mathrm{Tr}(\rho^{q})}{q - 1}.
\end{equation}
reparametrizes the same information content as R\'{e}nyi entropies but highlights
departures from extensivity in complex many-body systems. While black hole
replica calculations are formulated in terms of R\'{e}nyi entropies, one may view
$q$-deformations as a phenomenological tool to discuss effective non-factorizing
correlations. In this perspective the parameter $q$ labels departures from
perfect factorization, in loose analogy with the gravitational situation where
replica wormholes signal a breakdown of naive Hilbert-space factorization. This correspondence is heuristic and does not assert that the gravitational path
integral produces a Tsallis functional form. Rather, it suggests that both
replica wormholes in gravity and non-extensive entropies in statistical physics
reflect the broader theme that information-theoretic additivity can fail in
systems with strong correlations or geometric connectivity.

In the gravitational context the von Neumann entropy appears at the endpoint of
the replica construction, where the R\'enyi entropy is analytically continued to
$n \to 1$. In this sense $S_{\mathrm{vN}}$ may be viewed as the analogue of a
tangent value to the $n$ dependent information family. The computation of
$\mathrm{Tr}\,\rho^{n}$ in the gravitational path integral is implemented by the
replica trick and receives contributions from two distinct saddles: the Hawking
saddle and the replica wormhole saddle. The latter introduces correlations among
different replicas and accounts for the late time Page behavior. Formally the replica method introduces $n$ copies of the system. This raises an
apparent tension with the quantum no cloning theorem, which forbids the
duplication of an unknown quantum state. In the replica path integral however
the replicas are not physical copies prepared in the laboratory but rather
mathematical replicas in the sense of statistical field theory. The procedure
assumes an ensemble representation of the state rather than a physical
cloning process and therefore remains compatible with the no cloning theorem.
The gravitational saddle associated with replica wormholes can be interpreted as
a non factorizing contribution that reorganizes information without literally
replicating quantum states. The key lesson is that the parameter $n$ carries physical significance in
gravitational systems precisely because replica wormholes induce correlations
across copies. This motivates studying information theoretic quantities that
encode how the system evolves with $n$ and identifying an appropriate measure
for the change of entanglement during evaporation. Understanding this flow
provides a route to quantifying irreversibility in black hole radiation beyond
unitarity alone.

The replica index $n$ plays a distinguished role in gravity, since replica
wormholes introduce correlations across copies and thereby break naive
factorization. In parallel, non additive generalizations of entropy in
statistical mechanics introduce a deformation parameter $q$ that quantifies
departures from extensivity in the probability space~\cite{tsallis}. Although $n$ and $q$
arise in different constructions, they share a common conceptual meaning: both
parametrize the strength of correlations that obstruct a direct product
structure of subsystems. In this sense they may be regarded as different
coordinates on a single information theoretic axis measuring the degree of
non factorization. When $n \to 1$ or $q \to 1$ one recovers the additive,
factorizable limit appropriate to ordinary quantum field theory, whereas
$n \neq 1$ or $q \neq 1$ describe regimes where correlated configurations
contribute non trivially ~\cite{tsallis,tsallis1}.

We emphasize that this correspondence is conceptual and does not imply that the
gravitational path integral produces a Tsallis entropy. Rather, $n$ and $q$
provide complementary ways to quantify correlation induced departures from
additivity: $n$ in the geometric replica sector and $q$ in the probabilistic
sector. This suggests that effective $q$ deformations may serve as a useful
phenomenological language for describing non factorizing sectors that appear in
gravitational settings, without replacing the fundamental replica approach based
on R\'enyi entropies.

The structure and arrangement of this paper is as follows. Section 2 gives a review of the path integral of two-dimensional gravity to show how to obtain a "Hawking saddle" at early times of black hole evaporation and a "replica wormhole saddle" at late times. This may explain why the unitarity of quantum mechanics is not violated during black hole evaporation. Section 3 will explore in depth the relationship between the replica trick and the quantum no-cloning theorem. Through detailed analysis, it is revealed how the replica trick seems to challenge the basic principle of the quantum no-cloning theorem. Specifically, there is a reversibility problem: one can reproduce the full information of a known quantum state from n copies of the state. However, for an unknown quantum state, due to the limitation of the quantum no-cloning theorem, it is impossible to copy it, so other information of the quantum state cannot be measured. Section 4 focuses on modular entropy and entanglement capacity. By analogy with entropy in thermodynamics and acceleration in mechanics, it is pointed out that the relative entropy of Hawking saddle and replica wormhole saddle can be used as the key physical quantity to describe the irreversibility of the system, and the similarity between R\'{e}nyi entropy and thermodynamic free energy is discussed. In order to describe the change of state in the process of black hole evaporation, the relative entropy and the generalized second law are studied in Section 5, which shows that the evolution of the system is indeed irreversible. A summary and a discussion are given in the last section.

\section{Replica wormhole and wormhole saddle point}

\quad By reviewing the path integral of two-dimensional gravity, this paper explains how to obtain the ``Hawking saddle'' in the early stage of black hole evaporation and how to obtain the ``replica wormhole saddle'' in the late stage, so as to explain why the unitarity of quantum mechanics is not violated in the process of black hole evaporation. In the whole process of calculation, the two-dimensional Jackiw-Teitelboim (JT) gravity is coupled to a two-dimensional conformal field, which acts as a heat bath, grows in Minkowski space without gravity, and is connected to the gravitational field by a transparent boundary. See Refs.~\cite{35Penington2022,36Almheiri2020} for a detailed discussion.

Consider the action in two-dimensional gravity

\begin{align}
I &= I_{\text{JT}} + \mu \int_{\text{brane}} ds , \label{eq:I_total} \\[6pt]
I_{\text{JT}}
&= - \frac{S_0}{2\pi}
\left[
    \frac{1}{2}\int_{\mathcal{M}} \sqrt{g}\,R
    + \int_{\partial\mathcal{M}} \sqrt{h}\,K
\right]
-
\left[
    \frac{1}{2}\int_{\mathcal{M}} \sqrt{g}\,\phi(R+2)
    + \int_{\partial\mathcal{M}} \sqrt{h}\,\phi\,K
\right]. \label{eq:I_JT}
\end{align}

The second term in Eq.~(1) represents the contribution from the end-of-the-world (EOW) brane.
In Eq.~(2), $S_0 = 2\pi \phi_h$ is a constant determined by the dilaton value at the
horizon, $R$ is the Ricci scalar, $K$ is the extrinsic curvature, and $\phi$ denotes the
dilaton field, which supplies the degrees of freedom for two-dimensional gravity.

On the EOW brane, $k$ internal states are introduced. These states serve as the
entanglement partners of the black hole interior modes appearing in the early Hawking
radiation. The $k$ states are taken to be maximally entangled with an auxiliary reference
system $\mathcal{R}$. In this way, the early radiation of an evaporating black hole can be
represented explicitly. Treating the black hole and the auxiliary system $\mathcal{R}$ as
a combined quantum system, the total state may be written as
\begin{equation}
|\Psi\rangle = \frac{1}{\sqrt{k}} \sum_{i=1}^{k}
|\psi_i\rangle_{\mathrm{B}} \, |i\rangle_{\mathrm{R}} \, .
\label{eq:Psi_def}
\end{equation}
Here $|\psi_i\rangle_{\mathrm{B}}$ denotes the quantum state of the black hole, and the index $i$
indicates that the EOW brane is prepared in the state $i$. The state $|i\rangle_{\mathrm{R}}$
represents the quantum state of the auxiliary radiation system $\mathcal{R}$. The entropy
of the system $\mathcal{R}$ can be computed by the two-dimensional gravitational path
integral. The corresponding density matrix $\rho_{\mathcal{R}}$ takes the form
\begin{equation}
\rho_{\mathcal{R}}
= \frac{1}{k} \sum_{i,j=1}^{k}
|j\rangle_{\mathcal{R}}\langle i|_{\mathcal{R}}\,
\langle \psi_i|\psi_j\rangle_{\mathrm{B}} \, .
\label{eq:rho_R}
\end{equation}
The matrix element of the density matrix $\rho_{\mathcal{R}}$ is given by the
gravitational amplitude $\langle \psi_i|\psi_j\rangle_{\mathrm{B}}$, which is computed
via the two-dimensional gravitational path integral with appropriate boundary
conditions. Schematically, we write
\begin{equation}
\langle i|\rho_{\mathcal{R}}|j\rangle = Z_{ij}\, .
\end{equation}

\begin{figure}[htp]
    \centering
    \includegraphics[scale=0.35]{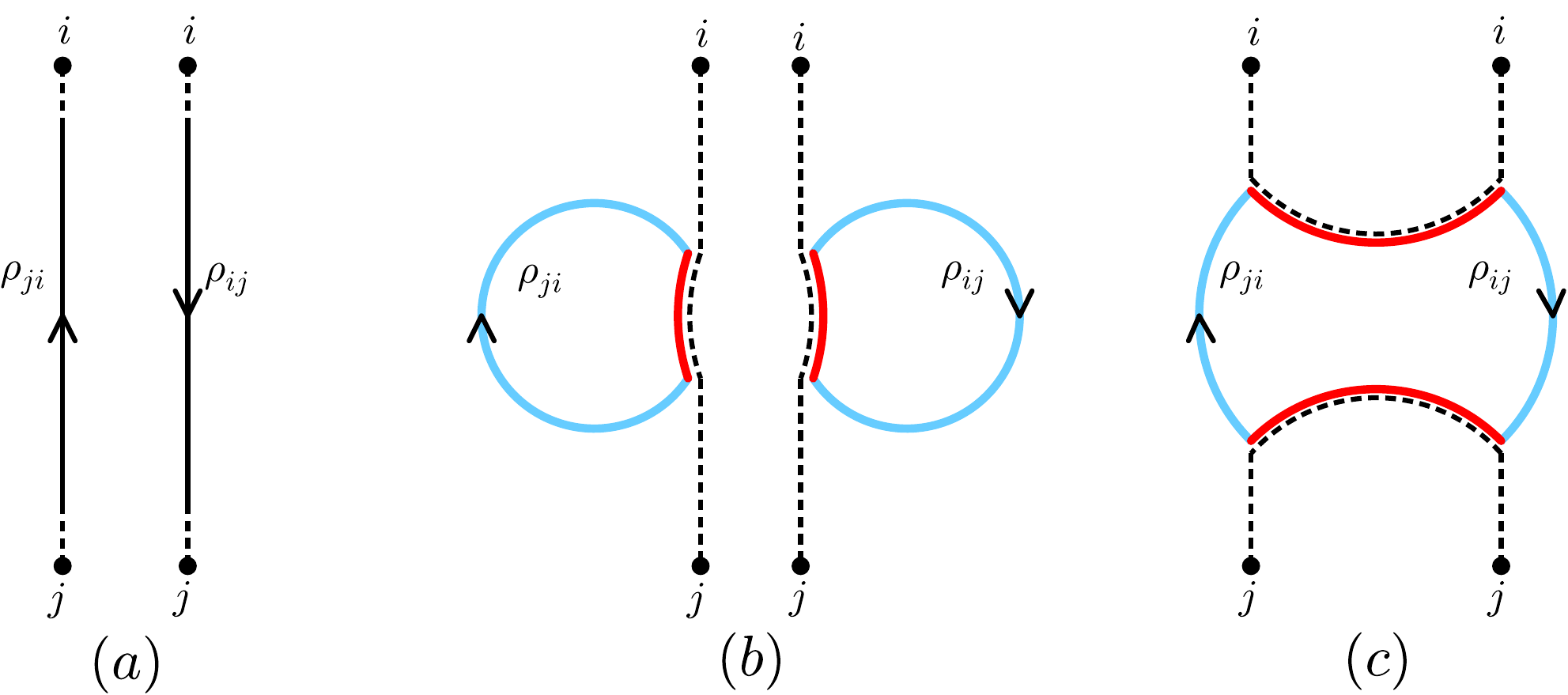}
    \caption{Schematic diagram for the gravitational path integral. The indices $i$ and $j$ represent different states. The gravitational region is shown in blue and the EoW brane is shown in red: (a) The bodunary conditions for the purity \eqref{eq6}; (b) and (c) are two different filling ways.}
    \label{fig:pathReplica}
\end{figure}

The purity (second R\'enyi entropy) can then be expressed in terms of these
matrix elements as
\begin{equation}
\mathrm{Tr}\,\rho_{\mathcal{R}}^{\,2}
= \sum_{i,j}
\langle i|\rho_{\mathcal{R}}|j\rangle
\langle j|\rho_{\mathcal{R}}|i\rangle
= \sum_{i,j} Z_{ij} Z_{ji} \, .
\label{eq6}
\end{equation}
The boundary condition for calculating the pruity \eqref{eq6} is shown in figure \ref{fig:pathReplica}.
Unlike the calculation of the von Neumann entropy, in which the indices $i$ and $j$ can be summed over through two equivalent approaches by connecting the dotted lines. Here the case involves two distinct topological configurations. The first type corresponds to a disconnected geometric configuration comprising two components with disk topology. While the second type features a connected structure, specifically, the Euclidean wormhole with disk topology. In the first case, the overall topology consists of two separate disks. However, for the later case, the geometry is linked through a pair of end of the world (EoW) branes that are joined together, resulting in a cylindrical topology.

Equivalently, using the gravitational partition functions on specific topologies one finds two types of saddle contributions (disconnected and connected) \cite{35Penington2022,36Almheiri2020}
\begin{equation}
\mathrm{Tr}\,\rho_R^2\;=\;\frac{Z_{\mathrm{disk}}^{(i)}Z_{\mathrm{disk}}^{(j)}+Z_{\mathrm{cyl}}^{(ij)}}{Z_{\mathrm{norm}}^2}\approx\frac{k\,Z_{\mathrm{disk}}^2+Z_{\mathrm{cyl}}}{Z_1^2}\,.
\end{equation}
If we extend to the general case and neglect contributions from higher topology
saddles, the trace of the $n$-th power of the density matrix can be written as
\[
\mathrm{Tr}\!\left(\rho_{\mathcal{R}}^{\,n}\right)
=
\frac{
k Z_1^{\,n}
+ k^2 \binom{n}{2} Z_2 Z_1^{\,n-2}
+ \cdots
+ k^i \binom{n}{i} Z_i Z_1^{\,n-i}
+ \cdots
+ k^n Z_n
}{
\left(k Z_1\right)^n
}\, .
\]

In JT gravity, an approximate expression for the partition function $Z_n$ can be obtained
by keeping only the contribution from the topological term $S_0$. In general, the
dependence of the partition function on the topological term takes the form
\(\mathrm{e}^{\chi S_0}\), where $\chi$ denotes the Euler characteristic. For generic $n$,
the geometries contributing to $Z_n$ correspond to disk topologies with $\chi = 1$,
so that
\[
Z_n \propto \mathrm{e}^{S_0} \, .
\]
Equation~(9) then yields
\begin{equation}
\mathrm{Tr}\!\left(\rho_{\mathcal{R}}^{\,2}\right)
= k^{-1} + \mathrm{e}^{-S_0}\, .
\label{eq:purity}
\end{equation}

If $k$ is small and the disconnected saddle dominates, the purity is
$\mathrm{Tr}(\rho_{\mathcal{R}}^{\,2}) \simeq 1/k$. When $k$ becomes large, the connected
saddle dominates, yielding $\mathrm{Tr}(\rho_{\mathcal{R}}^{\,2}) \simeq \mathrm{e}^{-S_0}$,
independent of $k$. The transition in dominance between these saddles is the mechanism
that prevents the radiation entropy, the R\'enyi entropy, from diverging. Since the total state is pure, this implies that the entropy of the excited EOW brane sector of the black hole remains finite even when $k$ is large. The
underlying mechanism is the weak non-orthogonality among the EOW states. This
accumulated non orthogonality becomes significant when
\(k \sim \mathrm{e}^{S_{\mathrm{BH}}}\), where \(S_{\mathrm{BH}}\) denotes the
Bekenstein Hawking entropy of the black hole.

Through the replica technique, we find that the quantity $\mathrm{Tr} \rho^2$ obtained from the gravitational path integral is influenced by non perturbative effects. These effects originate from new saddle point configurations known as replica wormholes in the gravitational path integral. The calculation involves introducing $n$ copies of the original black hole, analytically continuing the result to non-integer $n$, and then evaluating the von Neumann entropy. Generalizing to the general case, one can consider two extreme limits for arbitrarily large $n$:

1) The case of a fully unconnected geometry, whose topology is dominated by $n$ disks; in this case $k \ll e^{S_{\rm BH}}$. This case only has $k$ independent circle contributions, and one can write the partition function as
\begin{equation}
\mathrm{Tr}(\rho_R^n) \propto \frac{1}{k^{\,n-1}}\, .
\label{eq:unconnected}
\end{equation}

2) The case of a fully connected geometry, whose topology is dominated by a single disk; here $k \gg e^{S_{\rm BH}}$, in which case the numerator of the partition function is mainly contributed by $Z_n$:
\begin{equation}
\mathrm{Tr}(\rho_R^n)
\propto \frac{k^n Z_n}{k^n Z_1^n}
= \frac{Z_n}{Z_1^n}\, .
\label{eq:connected}
\end{equation}

The contributions of the $k$ cancel each other, so the trace of the density matrix is mainly contributed by the gravitational effect. To calculate the von Neumann entropy, one needs to expand $Z_n$ near $n=1$. One technique to achieve this continuation is to note that the geometry associated with $Z_n$ has a $\mathbb{Z}_n$ replica symmetry. When $n \to 1$, it reduces to the original unduplicated geometry, and the calculation of the von Neumann entropy reduces to
\begin{equation}
S = S_0 + 2\pi \phi_h\, ,
\end{equation}
where $\phi_h$ is the value of the dilaton field on the horizon.

In the early stage of black hole evaporation, the spacetime geometry contains $n$ independent copies of the original black hole. This configuration corresponds to the Hawking saddle of the von Neumann entropy. At the Hawking saddle, the von Neumann entropy increases linearly with time, which leads to the well-known tension between the semiclassical prediction for the radiation entropy and the unitarity requirement of quantum mechanics. In contrast, the replica wormhole saddle forms a new geometry by connecting the different copies, thereby reducing the entropy.

Up to this point, we have summarized the gravitational path integral in two-dimensional dilaton gravity, and we have clarified the circumstances under which Hawking saddles and replica wormhole saddles arise. The existence of both saddles ensures that the evolution of black holes is ultimately unitary, resolving the information paradox within this framework. It should be emphasized, however, that although explicit computations are currently only feasible in two-dimensional gravity, the physical implications are instructive. The subsequent discussion focuses on an additional conceptual layer, namely the directionality or invertibility of information flow in the moduli space during the transition from the Hawking saddle to the replica wormhole saddle.\footnote{See, for example, recent discussions on the microscopic reversibility of replica wormhole transitions \cite{modularentropy}.} Existing analyses indicate that the transition between the Hawking saddle and the replica wormhole saddle is microscopically unitary and thus reversible. According to the gravitational path integral, the early-time evaporation regime is dominated by the Hawking saddle, whereas the late-time regime is governed by the wormhole saddle, with the crossover occurring near the Page time. Throughout the evaporation process, unitarity is preserved and no information is lost. Nevertheless, a conceptual puzzle remains: why does the Hawking saddle dominate at early times, instead of the replica wormhole saddle?

In replica calculations, one formally introduces $n$ copies of the density matrix. However, the quantum no-cloning theorem forbids the cloning of an unknown quantum state. Therefore, to prepare $n$ replicas of the density matrix, the underlying state $\lvert \Psi \rangle$ must be known in advance. In other words, in order to emulate black hole evaporation experimentally, one must prepare $n$ identical and known quantum states at the outset. A finite von Neumann entropy then emerges by analytically continuing to the limit $n \to 1$. From an informational perspective, this suggests a directional constraint: only when the relevant information of the quantum state $\lvert \Psi \rangle$ is available can a replica wormhole saddle arise in the subsequent gravitational evolution. Consequently, the consistency with the quantum no-cloning theorem implies that, conceptually, the multi-copy Hawking radiation configuration should be regarded as existing \emph{prior} to the single-copy configuration in the evaporation process. The existence of multiple replicas prior to analytical continuation suggests
that the relevant Hilbert space structure is not strictly factorized during
black hole evaporation. If replicas were entirely independent, each copy
would evolve autonomously and no information could be recovered from the
radiation. The emergence of replica wormholes precisely signals a departure
from naive factorization: different replicas become coupled in the
gravitational path integral, enabling correlations that restore unitarity.

In this sense, replica wormholes reflect a non-factorized Hilbert space structure in quantum gravity. When gravitational saddles remain disconnected, the von Neumann entropy of each replica behaves additively, reproducing Hawking's result. When wormhole saddles contribute, correlations between replicas emerge, and the flow of information between replicas becomes possible, leading to the Page curve.
From this perspective, Tsallis-type non-additive entropies reflect generalized correlations in the probability distributions of composite systems, going beyond standard additive measures, whereas replica wormholes encode correlations between different replicas of the gravitational system. The physical universe is therefore not strictly factorizable: wormholes provide a geometric realization of non-factorization, while Tsallis statistics provide a probabilistic realization of non-factorization~\cite{tsallis1}. In summary, replica wormholes represent the geometric manifestation of a non-factorized Hilbert space in quantum gravity, just as Tsallis statistics represent non-additivity in quantum probability theory.

\section{Replica trick and quantum no-cloning theorem}

The replica trick is a standard method for treating disordered systems and complex interactions in statistical physics. Its essential idea is to introduce multiple replicas of a system and extract global information by averaging over them. For a quantum system with density matrix $\rho$, one constructs $n$ copies and considers $\mathrm{Tr}(\rho^{n})$. The free energy can then be obtained from
\begin{equation}
F = - \frac{1}{\beta} \lim_{n \to 0} \frac{\partial}{\partial n} \ln \left\langle \mathrm{Tr}(\rho^{n}) \right\rangle .
\label{eq:replica_free_energy}
\end{equation}
In this procedure, one does not attempt to physically clone an unknown quantum state $\lvert \psi \rangle$. Rather, one treats multiple mathematical replicas of the system to deduce macroscopic properties.

The quantum no-cloning theorem states that it is impossible to copy an unknown quantum state \cite{37Wootters1982}. Consider an arbitrary state
\begin{equation}
\lvert \psi \rangle = \alpha \lvert \phi_{1} \rangle + \beta \lvert \phi_{2} \rangle ,
\label{eq:unknownstate}
\end{equation}
where $\alpha$ and $\beta$ are unknown complex coefficients. The theorem asserts that no quantum operation $U$ can map
\[
\lvert \psi \rangle \otimes \lvert 0 \rangle ~\longrightarrow~ \lvert \psi \rangle \otimes \lvert \psi \rangle
\]
for every possible $\lvert \psi \rangle$. The proof proceeds by contradiction. Assume that there exists a unitary operator $U$ satisfying
\[
U \lvert \psi \rangle = \lvert \psi \rangle \otimes \lvert \psi \rangle ,
\qquad
U \lvert \phi \rangle = \lvert \phi \rangle \otimes \lvert \phi \rangle ,
\]
for two arbitrary pure states $\lvert \psi \rangle$ and $\lvert \phi \rangle$. Consider the inner product between the two states before and after applying $U$:
\begin{equation}
\langle \psi \otimes 0 \lvert U^{\dagger} U \rvert \phi \otimes 0 \rangle
=
\langle \psi \otimes \psi \lvert \phi \otimes \phi \rangle .
\label{eq:innerproduct}
\end{equation}
Using the unitarity condition $U^{\dagger} U = I$, we obtain
\begin{equation}
\langle \psi \rvert \phi \rangle
=
\langle \psi \otimes \psi \rvert \phi \otimes \phi \rangle .
\label{eq:innerproduct2}
\end{equation}
The left-hand side equals
$
\langle \psi \rvert \phi \rangle ,
$
while the right-hand side equals
$
\bigl\lvert \langle \psi \rvert \phi \rangle \bigr\rvert^{2}.
$
Let $\langle \psi \vert \phi \rangle = c$. The above equality implies
\[
c = c^{2}.
\]
The solutions are $c = 0$ or $c = 1$. For two distinct quantum states, the overlap should satisfy $0 < \lvert c \rvert < 1$, which contradicts the above identity. The assumption is therefore false, and an unknown quantum state cannot be cloned.

The scope of applicability of the quantum no-cloning theorem can be further clarified from the perspective of the replica method. Consider a quantum system whose state is described by a density matrix $\rho$. Suppose one introduces multiple replicas $\lvert \psi_i \rangle$, together with a random variable $R_{ij}$ that characterizes a small disturbance between different quantum states:
\begin{equation}
\langle \psi_i \vert \psi_j \rangle = c_{ij} + e^{-S_{0}/2} R_{ij} ,
\label{eq:randomoverlap}
\end{equation}
where $S_{0}$ denotes the entropy of the black hole. Typically $S_{0}$ is very large, which makes $e^{-S_{0}/2}$ exponentially suppressed, and $R_{ij}$ has zero mean and finite variance.

Let $\lvert \psi_i \rangle$ and $\lvert \psi_j \rangle$ denote two distinct quantum states. Suppose a unitary operation $U$ exists that maps an input state $\lvert \psi_i \rangle \otimes \lvert 0 \rangle$ to $\lvert \psi_i \rangle \otimes \lvert \psi_i \rangle$. In contrast to the standard argument, here one must also take the ensemble average over the random variables $R_{ij}$. Let the ensemble averages satisfy $\overline{c_{ij}} = c$, $\overline{R_{ij}} = 0$, and $\overline{R_{ij}^{2}} = \sigma^{2}$, where an overline denotes averaging. Then
\begin{equation}
\overline{\langle \psi_i \otimes 0 \vert U^{\dagger} U \vert \phi_j \otimes 0 \rangle}
=
\overline{\langle \psi_i \otimes \psi_i \vert \phi_j \otimes \phi_j \rangle} .
\label{eq:ensembleunitarity}
\end{equation}
This yields
\begin{equation}
\overline{\langle \psi_i \vert \phi_j \rangle}
=
\overline{\lvert \langle \psi_i \vert \phi_j \rangle \rvert^{2}} ,
\label{eq:ensembleinnerproduct}
\end{equation}
or equivalently
\begin{equation}
c = c^{2} + e^{-S_{0}/2} \sigma^{2}.
\label{eq:solution_c}
\end{equation}
In this expression, $c$ has a nontrivial solution within the interval $(0,1)$, so the usual no-cloning argument can fail. Thus, when correlations among replicas are allowed, the replica construction may in principle lead to a violation of the quantum no-cloning theorem. To avoid such situations, the complete information of the quantum state must be known before any replication procedure is applied.

In the replica method, one considers multiple copies of a known system rather than attempting to copy an unknown quantum state directly. Each replica is an independent system whose state is known, and although interactions between replicas may be introduced, they occur at the level of the path integral rather than physical state duplication. Even though an unknown quantum state cannot be cloned, global information can be extracted by statistical analysis of multiple known replicas. The no-cloning theorem restricts operations that attempt to duplicate an unknown state, whereas the replica method bypasses this restriction by working with averaged quantities over replicas instead of producing physical copies of unknown quantum states. The replica technique relies on mathematical tools, such as path integrals, to access averaged observables without violating quantum mechanics.

This consideration highlights a reversibility condition. From $n$ copies of a known quantum state, one may reconstruct all of its information. However, one cannot access otherwise unknown information about a quantum state by attempting to generate additional copies of it.

\section{Modular entropy and entanglement capacity}

\quad In the above discussion, Zeno's paradox was compared with the information loss problem of black holes. Taking the limit of the R\'{e}nyi entropy is analogous to taking the limit in the definition of instantaneous velocity, where the displacement $\Delta x$ is divided by the time interval $\Delta t$ and subsequently the limit $\Delta t \to 0$ is taken. In this manner, the von Neumann entropy acquires a more rigorous mathematical status, since it can be regarded as a conserved quantity in a manner analogous to conserved quantities in classical mechanics. At both the beginning and the end of Hawking radiation, the von Neumann entropy of the Hawking radiation remains unchanged. The recent derivations of the Page curve for black hole evaporation confirm this property.

However, this observation alone is insufficient. A self-consistent physical theory requires, in addition to invariants, well-defined quantities that characterize dynamical change. In classical mechanics, change is characterized by acceleration, while in thermodynamics, change is characterized by entropy. The physical quantity conjugate to acceleration is force, and entropy encodes the irreversibility of a process. It is important to note that thermodynamic quantities cannot be captured simply by the mathematical structure of analytical mechanics. Thermodynamics possesses a more universal conceptual framework. As will be seen below, the mathematical formulation of entanglement entropy is closely aligned with the theoretical structure of thermodynamics.

To develop a systematic description of modular entropy and motivate the introduction of new physical quantities, an intuitive approach is to draw an analogy. In what follows, we compare entanglement entropy with statistical mechanics and explain why the replica parameter may be interpreted as an effective inverse temperature. For a quantum field theory, the von Neumann entanglement entropy $S_A$ associated with a subregion $\mathcal{H}_A$ is defined as

\begin{equation}
S_{\mathrm{vN}}(\rho_A) = - \mathrm{Tr}\!\left( \rho_A \log \rho_A \right),
\label{eq:vNentropy}
\end{equation}
where the reduced density matrix $\rho_A$ is given by
\begin{equation}
\rho_A = \mathrm{Tr}_{\bar{A}} \left( \rho_{\mathrm{total}} \right).
\label{eq:reduceddensity}
\end{equation}
Here $\bar{A}$ denotes the complement of the subsystem $A$. The total density matrix $\rho_{\mathrm{total}}$ is taken to be a pure state, satisfying $\mathrm{Tr}\!\left(\rho_{\mathrm{total}}\right)=1$.

The von Neumann entropy quantifies the degree of mixedness of a quantum state and measures the number of maximally entangled pairs required to represent the state. It vanishes for a pure state and reaches its maximal value for a maximally mixed state, thus characterizing the distinguishability of quantum states. If $\rho_{\mathrm{total}}$ is mixed, then $\mathrm{Tr}(\rho_{\mathrm{total}}^{\,n}) < 1$ for $n>1$.

The entanglement entropy satisfies an additivity (or strong subadditivity) relation
\begin{equation}
S_{AC} + S_{CB} \geq S_C + S_{ABC}, \qquad
S_A + S_B \geq S_C + S_{AB} .
\label{eq:SSA}
\end{equation}
More generally, one may introduce the R\'enyi entropy~\cite{Renyi1961}, \begin{equation}
S_{n}(\rho) = - \frac{1}{n-1} \log \mathrm{Tr}\!\left(\rho^{\,n}\right).
\label{eq:RenyiEntropy}
\end{equation}
The von Neumann entropy can be obtained from the R\'enyi entropy by taking the limit $n \to 1$,
\begin{equation}
S_{\mathrm{vN}}(\rho)
= - \lim_{n \to 1} \frac{1}{n-1} \log \mathrm{Tr}\!\left( \rho^{\,n} \right).
\label{eq:RenyiToVN}
\end{equation}
Later we will see that the R\'enyi entropy takes a form analogous to the free energy in thermodynamics. The R\'enyi entropy satisfies the following inequalities~\cite{Renyi1961}:
\begin{equation}
\frac{\partial S_n}{\partial n} \le 0,
\qquad
\frac{\partial}{\partial n} \left( \frac{n-1}{n} S_n \right) \ge 0,
\qquad
\frac{\partial^{2}}{\partial n^{2}} \bigl[ (n-1) S_n \bigr] \ge 0.
\label{eq:RenyiInequalities}
\end{equation}
We now introduce the modular Hamiltonian
\begin{equation}
H_A = - \log \rho_A.
\label{eq:ModularHamiltonian}
\end{equation}
In terms of $H_A$, the R\'enyi entropy can be expressed as
\begin{equation}
S_n(\rho)
= \frac{1}{1-n} \log \mathrm{Tr}_A \!\left( e^{-n H_A} \right).
\label{eq:RenyiModular}
\end{equation}
For comparison, recall the definition of the free energy in statistical mechanics,
\begin{equation}
F = - \frac{1}{\beta} \log \mathrm{Tr}\!\left( e^{-\beta H} \right).
\label{eq:FreeEnergy}
\end{equation}
The formal similarity between Eq.~\eqref{eq:RenyiModular} and Eq.~\eqref{eq:FreeEnergy} is manifest: the replica parameter $n$ plays the role of an effective inverse temperature $\beta$. In the following, we discuss the interpretation of $n$ and the physical meaning of $S_m$.

In contrast to the replica trick and conventional statistical mechanics, we introduce two additional physical quantities in this work.

\begin{enumerate}
\item \textbf{Modular entropy.}
Modular entropy is a quantity that arises in quantum information theory, defined in Table~1.
Unlike the von Neumann entropy, the modular entropy depends on the replica parameter $n$ through a derivative with respect to $n$, and provides a refined measure of the structure of the entanglement within the system.
The motivation for introducing modular entropy originates from studies of the holographic dual of the R\'enyi entropy. In particular, it was realized that one may define an alternative entropy functional that behaves more analogously to a ``thermodynamic entropy''. Within the holographic framework, the R\'enyi entropy is dual to the area of a cosmic brane inserted in the gravitational bulk geometry~\cite{Dong2016}.

\item \textbf{Entanglement capacity.}
The entanglement capacity is another key quantity, defined as the change of modular entropy when the replica number is varied. Characterizes the response of the system to adding or removing a replica and therefore serves as an indicator of possible phase transitions in the modular space.
\end{enumerate}

\begin{table}[!t]
\centering
\caption{Analogy between statistical mechanical quantities and modular quantities}
\renewcommand{\arraystretch}{1.25}
\scriptsize
\begin{tabularx}{\textwidth}{l X l X}
\toprule
\textbf{Statistical mechanical quantity} & \textbf{Expression}
& \textbf{Modular analogy} & \textbf{Expression} \\
\midrule

Inverse temperature
& $\beta$
& Replica parameter
& $n$ \\

Hamiltonian
& $H$
& Modular Hamiltonian
& $H = -\log \rho_A$ \\

Partition function
& $Z(\beta)=\mathrm{Tr}(e^{-\beta H})$
& Replica partition function
& $Z_n(n)=\mathrm{Tr}(e^{-nH_A})$ \\

Free energy
& $F(\beta)= -\frac{1}{\beta}\log \mathrm{Tr}(e^{-\beta H})$
& Replica free energy
& $F(n)= -\frac{1}{n}\log[\mathrm{Tr}(\rho_A^n)]$ \\

Energy
& $E(\beta)= -\partial_{\beta}\log[\mathrm{Tr}(e^{-\beta H})]$
& Replica energy
& $E(n)= -\partial_{n}\log[\mathrm{Tr}(\rho_A^n)]$ \\

Thermodynamic entropy
& $\begin{aligned}
S(\beta) &= \log[\mathrm{Tr}(e^{-\beta H})] \\
&\quad - \beta \partial_\beta \log[\mathrm{Tr}(e^{-\beta H})]
\end{aligned}$
& Mode entropy
& $\begin{aligned}
S_m &= \log[\mathrm{Tr}(\rho_A^n)] \\
&\quad - n\partial_n \log[\mathrm{Tr}(\rho_A^n)]
\end{aligned}$ \\

Heat capacity
& $C(\beta)=\beta^2\partial_{\beta}^2\log[\mathrm{Tr}(e^{-\beta H})]$
& Entanglement capacity
& $C_n = n^2 \partial_{n}^2\log[\mathrm{Tr}(\rho_A^n)]$ \\

\bottomrule
\end{tabularx}
\end{table}
Next, we elaborate on the analogy between the R\'enyi entropy and thermodynamic quantities.
The modular entropy is defined as
\begin{equation}
S_m = \frac{1}{n^{2}} \, \partial_{n} \!\left( \frac{n-1}{n} S_n \right).
\label{eq:Sm_def}
\end{equation}
If the replica parameter $n$ is interpreted as an effective inverse temperature, this definition becomes structurally analogous to the thermodynamic entropy. In particular, the modular entropy can be expressed as
\begin{equation}
S_m(\rho)
= - n^2 \, \partial_{n} \!\left[ \frac{1}{n}
\log \mathrm{Tr}_{A}\left(e^{-n H_A}\right) \right],
\label{eq:Sm_z}
\end{equation}
where $H_A = -\log \rho_A$ denotes the modular Hamiltonian. In comparison, the thermodynamic entropy can be written in the form
\begin{equation}
S = - \beta^{2} \,
\partial_{\beta} \!\left[ \frac{1}{\beta}
\log \mathrm{Tr}\!\left(e^{-\beta H}\right) \right].
\label{eq:thermal_entropy}
\end{equation}
Identifying the replica parameter $n$ with the inverse temperature $\beta$, the formal equivalence between \eqref{eq:Sm_z} and \eqref{eq:thermal_entropy} becomes manifest. As summarized in Table~1, the modular Hamiltonian $H_A$ plays the role of the Hamiltonian in statistical mechanics, and the replica index $n$ plays the role of an effective inverse temperature. However, it should be emphasized that $n$ is not a physical temperature but rather counts the number of replicated Hilbert spaces in the replica construction. Within this correspondence, quantities such as the partition function, free energy, internal energy, modular entropy, and entanglement capacity admit natural counterparts in the moduli space. A more complete treatment of these relations will be presented in a forthcoming work~\cite{tsallis}.

As an illustrative example, consider the Hamiltonian
\begin{equation}
H = - \hbar \omega \, \sigma_{z},
\label{eq:Hamiltonian}
\end{equation}
where $\sigma_{z}$ denotes the Pauli matrix,
\[
\sigma_{z} =
\begin{pmatrix}
1 & 0 \\
0 & -1
\end{pmatrix}.
\]
This system describes a Bell state composed of two subsystems $A$ and $B$. The density matrix of the composite system at zero temperature is
\begin{equation}
\rho_{AB} = \frac{1}{2}
\begin{pmatrix}
1 & 0 & 0 & 1 \\
0 & 0 & 0 & 0 \\
0 & 0 & 0 & 0 \\
1 & 0 & 0 & 1
\end{pmatrix}.
\label{eq:rhoAB}
\tag{34}
\end{equation}
Tracing over subsystem $B$ yields the reduced density matrix for $A$,
\begin{equation}
\rho_{A} = \mathrm{Tr}_B(\rho_{AB})
= \frac{1}{2}
\begin{pmatrix}
1 & 0 \\
0 & 1
\end{pmatrix}.
\label{eq:rhoA_zeroT}
\end{equation}
At finite temperature, we assume that subsystems $A$ and $B$ are each in thermal equilibrium $\rho_{A} = \frac{e^{-\beta H_A}}{Z_A},~\rho_{B} = \frac{e^{-\beta H_B}}{Z_B}$.
If the Hamiltonians of the two subsystems are identical, then
\begin{equation}
\rho_A
=
\frac{1}{2 \cosh(\beta \hbar \omega)}
\begin{pmatrix}
e^{\beta \hbar \omega} & 0 \\
0 & e^{-\beta \hbar \omega}
\end{pmatrix}.
\label{eq:rhoA_finiteT}
\end{equation}
Consequently, we have
\begin{equation}
\rho_A^{n}
=
\left[\frac{1}{2 \cosh(\beta \hbar \omega)}\right]^{n}
\begin{pmatrix}
e^{n \beta \hbar \omega} & 0 \\
0 & e^{-n \beta \hbar \omega}
\end{pmatrix},
\label{eq:rhoA_n}
\end{equation}
and its trace is
\begin{equation}
\mathrm{Tr}(\rho_A^{n})
=
2 \left[\frac{1}{2 \cosh(\beta \hbar \omega)}\right]^{n}
\cosh(n \beta \hbar \omega).
\label{eq:Tr_rhoA_n}
\end{equation}
The modular entropy then takes the form
\begin{equation}
\begin{aligned}
S_m &=
n \log 2
- n \log\!\left[\mathrm{sech}(\beta \hbar \omega)\right] \\
&\quad + \log\!\left[2^{1-n} \cosh(n \beta \hbar \omega)
\mathrm{sech}^{\,n}(\beta \hbar \omega)\right]
- n \beta \hbar \omega \, \tanh(n \beta \hbar \omega).
\end{aligned}
\label{eq:Sm_Bell}
\end{equation}
The entanglement capacity is
\begin{equation}
C_n = n^{2} \partial_n^{2} \log\!\left( \mathrm{Tr}\rho_A^n \right)
= (n \beta \hbar \omega)^{2} \left[ 1 - \tanh^{2}(n \beta \hbar \omega) \right].
\label{eq:Cn}
\end{equation}
For fixed $\beta$, both the modular entropy and the entanglement capacity vary with the replica parameter $n$. In the limit $n \to 1$, the entanglement capacity is given by
\begin{equation}
C_1 = (\beta \hbar \omega)^{2} \left[1 - \tanh^{2}(\beta \hbar \omega)\right].
\label{eq:C1}
\end{equation}
As  $n \to \infty$, we find $C_{\infty} \to 0$, so that $C_1 > C_{\infty}$. In the zero-temperature limit ($\beta \to \infty$), $\tanh(n \beta \hbar \omega) \to 1$, and the entanglement capacity vanishes. As the temperature increases (smaller $\beta$), the entanglement capacity increases, reflecting the thermal enhancement of quantum fluctuations. For fixed $\beta$, increasing $n$ suppresses both $S_m$ and $C_n$, indicating a tendency toward an effectively ground state like  behavior in the high replica limit. In statistical mechanics, the de~Broglie wavelength $\lambda_{\mathrm{dB}} = \sqrt{\frac{\pi \hbar^2}{m k_B T}}$ distinguishes thermal and quantum regimes: when $\hbar \omega \ll 1$ thermal effects dominate, while for $\beta \hbar \omega \gg 1$ quantum effects prevail.

To characterize the role of the replica parameter, one may introduce an effective generalized de~Broglie wavelength
\begin{equation}
\lambda_{\mathrm{eff}} \equiv \lambda_{\mathrm{dB}} \, e^{S_{\mathrm{vN}} - S_m}.
\label{eq:lambda_eff}
\end{equation}
When $n \to \infty$, $\lambda_{\mathrm{eff}}$ increases, signaling enhances quantum effects induced by replica contributions. In contrast, for $n \to 1$ one has $S_{\mathrm{vN}} = S_m$, and the generalized wavelength reduces to the standard statistical-mechanical form.
The change of entropy in black hole radiation consists of two stages: an initial phase in which the entropy increases approximately linearly with time, followed by a phase in which the entropy gradually decreases. The transition point between these two regimes is the Page time $t_{\mathrm{Page}}$, which is parametrically long. In general, it can be estimated as~\cite{19Page2013}
\begin{equation}
t_{\mathrm{Page}} = \frac{6 S_{\mathrm{BH}}}{\kappa\, c},
\label{eq:tpage}
\end{equation}
where $S_{\mathrm{BH}}$ is the Bekenstein--Hawking entropy, $\kappa$ is the surface gravity, and $c$ is the central charge. The Page time is approximately half of the total black hole lifetime; for a stellar-mass black hole, this duration exceeds the current age of the Universe.

For massive black holes, evaporation proceeds extremely slowly and may be regarded as quasi-static, meaning the system remains arbitrarily close to equilibrium at each instant. Under this assumption, the evaporation process is approximately reversible and the radiation entropy is conserved; this yields the familiar Page curve, in which the entropy of the Hawking radiation first rises and then falls, consistent with unitary evolution at the microscopic level. In realistic physical situations, however, irreversibility is generally present. Thus, additional quantities are required to characterize irreversible effects. As discussed later, an $n$-dependent relative entropy serves as a suitable measure of such irreversibility in the replica framework.

\section{Relative entropy and generalized second law}

\quad According to the correspondence between the R\'enyi entropy and statistical-mechanical quantities, the behavior of the modular entropy exhibits a structure more closely analogous to thermodynamic entropy. In this context, it is instructive to recall the formulation of the generalized second law (GSL) in quantum field theory. In 2008, Casini~\cite{Casini2008} introduced a formulation of the GSL that avoids the conventional difficulties associated with defining entropy and energy in a local region in quantum field theory, such as nonlocality and ultraviolet divergences. The key idea is to subtract the vacuum contribution, defining the entropy difference
\begin{equation}
S_V = S(\rho_V) - S(\rho_V^{\,0}),
\end{equation}
and the corresponding energy difference
\begin{equation}
K_V = \mathrm{Tr}(K \rho_V) - \mathrm{Tr}(K \rho_V^{\,0}),
\end{equation}
where $K$ denotes the modular Hamiltonian associated with the vacuum state $\rho_V^{\,0}$.

Casini showed that the Bekenstein bound can be recast in terms of the relative entropy $S(\rho_V\|\rho_V^{\,0})$, a positive quantity measuring the statistical distinguishability between states. In the black-hole context, when an object crosses the horizon, the least constraining inequality is
\begin{equation}
S(\rho_1 \| \rho_{\mathrm{HH}}) \ge 0,
\end{equation}
where $\rho_1$ is the quantum state describing the exterior region at an initial time $t_1$, and $\rho_{\mathrm{HH}}$ is the Hartle--Hawking thermal state. This inequality implies that the distinguishability of any physically realizable state with respect to the thermal equilibrium state cannot decrease, which ensures the validity of the generalized second law. Even when matter falls into the black hole, the sum of the entropy outside the horizon and the Bekenstein--Hawking entropy does not decrease, preserving the total entropy budget and thereby respecting the GSL.

As discussed earlier, the Hawking saddle dominates in the initial stage of black hole evaporation, when the area of the minimal quantum extremal surface is negligible and replica wormhole contributions can be disregarded. At late times, the replica wormhole saddle becomes dominant. To make the physical role of the replica index $n$ more transparent, we now formulate an $n$-dependent version of the generalized second law in terms of relative entropy, thereby illustrating that $n$ carries physical significance rather than being merely a computational device.

For two density matrices $\rho$ and $\sigma$, the relative entropy is defined as
\begin{equation}
S(\rho\|\sigma) = \mathrm{Tr}(\rho \log \rho) - \mathrm{Tr}(\rho \log \sigma).
\label{eq:relative_entropy}
\end{equation}
Relative entropy obeys two key properties: positivity and monotonicity. Positivity states that $S(\rho\|\sigma) \ge 0$ for any $\rho$ and $\sigma$, with equality if and only if $\rho = \sigma$. The monotonicity property asserts that relative entropy cannot increase under any completely positive trace-preserving (CPTP) map. That is, for any CPTP map $\Phi$,
\begin{equation}
S(\Phi(\rho) \| \Phi(\sigma)) \le S(\rho\|\sigma).
\label{eq:monotonicity}
\end{equation}
Physically, monotonicity implies that no physically allowed quantum operation, such as unitary evolution, coarse-graining, or measurement, can increase our ability to distinguish $\rho$ from $\sigma$. In other words, information processing cannot make two quantum states more distinguishable than they initially were, consistent with the second law structure in quantum information theory.

We now consider $\sigma$ as a reference state and introduce its modular Hamiltonian $H_A$. One may then define the modular free energy
\begin{equation}
F(\rho) = \mathrm{Tr}(\rho H_A) - \frac{1}{n} S(\rho),
\end{equation}
where $S(\rho)$ denotes the von Neumann entropy of $\rho$. Using this definition, one finds
\begin{equation}
S(\rho\|\sigma) = n \left[ F(\rho) - F(\sigma) \right].
\label{eq:rel_free_energy}
\end{equation}

In the black hole evaporation process, let $\rho$ denote the density matrix associated with the Hawking saddle, and let $\rho_{n} = e^{-n H}$ correspond to the replica wormhole saddle. Comparing each of these with the Hartle--Hawking equilibrium state $\rho_{\mathrm{HH}}$, the monotonicity of relative entropy in Eq.~\eqref{eq:monotonicity} implies
\begin{equation}
S(\rho \| \rho_{\mathrm{HH}}) \le S(\rho_{n} \| \rho_{\mathrm{HH}}).
\label{eq:monotonicity_evap}
\end{equation}
Thus, as evaporation proceeds, the distinguishability (or information loss) of the physical state relative to the thermal Hartle--Hawking state does not increase. Using Eq.~\eqref{eq:rel_free_energy}, this inequality yields
\begin{equation}
F(\rho_{n}) - F(\rho) \le 0,
\end{equation}
which may be rewritten as
\begin{equation}
\left( \langle H_A \rangle_{n} - S_{n} \right) - \left( \langle H_A \rangle_{\rho} - S_{\rho} \right) \le 0.
\label{eq:free_energy_relation}
\end{equation}
Equivalently,
\begin{equation}
\left( S_{n} - S_{\rho} \right) - n \left( \langle H \rangle_{n} - \langle H \rangle_{\rho} \right) \ge 0,
\label{eq:final_directionality}
\end{equation}
where the subscript $\rho$ denotes the Hawking saddle, while the subscript $n$ refers to the replica wormhole saddle.

Relation~\eqref{eq:final_directionality} shows that there exists a directionality in the moduli space between the Hawking saddle and the replica wormhole saddle, governed by the second law of relative entropy. In this sense, the replica index $n$ plays the role of an effective inverse temperature. This argument parallels Casini's reformulation of the generalized second law in terms of relative entropy.

\section{Summary and discussion}

\quad In this work we reviewed recent progress on the black hole information problem and
formulated a unified viewpoint linking replica wormholes, non-additivity, and
information recovery. By drawing an analogy with Zeno's paradox, we emphasized
that resolving the information question requires addressing two layers: the
semi-classical Hawking saddle, and the quantum-gravitational replica wormhole
saddle responsible for restoring unitarity. The replica analysis shows that the
Hawking saddle governs early-time evaporation, while the replica wormhole saddle
dominates at late times, yielding the Page curve.

We further examined the quantum no-cloning theorem in relation to the replica
method, noting that consistency demands considering multiple copies of a known
quantum state rather than attempting to clone an unknown state. This clarifies
that replica copies must exist in principle from the outset of black hole
evolution, with negligible coupling at early times and dominant coupling at late
times, enabling information recovery. This viewpoint evokes a structure
reminiscent of many-worlds dynamics in quantum theory, where distinct sectors
become effectively coupled at the quantum-gravitational level.

To characterize the dynamics of entanglement during evaporation, we employed the
modular Hamiltonian and showed that modular entropy displays a thermodynamic form,
leading to a ``modular thermodynamics'' picture and the notion of entanglement
capacity. Our explicit computation for a Bell state demonstrates that both modular
entropy and entanglement capacity approach finite or vanishing limits as the replica
parameter $n$ increases, consistent with an interpretation of $n$ as an effective
inverse temperature.  To sharpen the physical meaning of $n$, we introduced an $n$-dependent formulation
of the generalized second law based on relative entropy. The positivity and
monotonicity of relative entropy ensure that the distinguishability between
replica wormhole saddles and the Hawking saddle evolves in a well-defined direction,
establishing a monotonic flow from the Hawking saddle to the wormhole saddle, and
confirming $n$ as a thermodynamic-like deformation parameter in the replica framework.

Finally, our analysis highlights a deeper organizing principle: replica wormholes
represent a breakdown of naive factorization in quantum gravity, corresponding to
non-additivity of Tsallis entanglement entropy, while Tsallis-type generalized entropies
offer a parallel probabilistic framework for non-additive correlations. In this
sense, wormholes provide a geometric realization of non-factorization, and Tsallis
statistics provide a statistical realization, together offering a coherent
interpretation of information preservation in quantum gravity.
\section*{Acknowledgements}

The author would like to thank Jianxin Lu, Sangjin Sin, Zhenbin Yang for helpful discussions.  Also, thanks Minghui Yu for useful discussions and comments on an earlier draft of this work. This work was partially supported by the National Natural Science Foundation of China under grant No. 12275166 and No. 12311540141.

\end{document}